\documentclass{article}

\usepackage{arxiv}

\usepackage[utf8]{inputenc} 
\usepackage[T1]{fontenc}    
\usepackage{hyperref}       
\usepackage{url}            
\usepackage{booktabs}       
\usepackage{amsfonts}       
\usepackage{nicefrac}       
\usepackage{microtype}      
\usepackage{xcolor}         

\usepackage[american]{babel}

\usepackage{amsmath}
\usepackage{amssymb}
\usepackage{amsthm}

\usepackage{mathtools}
\usepackage{ebproof}

\usepackage[all,cmtip]{xy}

\usepackage{txfonts}
\usepackage{pxfonts}

\makeatletter
\newcommand{\xdashrightarrow}[2][]{\ext@arrow 0359\rightarrowfill@@{#1}{#2}}
\newcommand*{\doublerightarrow}[2]{\mathrel{
  \settowidth{\@tempdima}{$\scriptstyle#1$}
  \settowidth{\@tempdimb}{$\scriptstyle#2$}
  \ifdim\@tempdimb>\@tempdima \@tempdima=\@tempdimb\fi
  \mathop{\vcenter{
    \offinterlineskip\ialign{\hbox to\dimexpr\@tempdima+1em{##}\cr
    \rightarrowfill\cr\noalign{\kern.5ex}
    \rightarrowfill\cr}}}\limits^{\!#1}_{\!#2}}}
\newcommand*{\triplerightarrow}[1]{\mathrel{
  \settowidth{\@tempdima}{$\scriptstyle#1$}
  \mathop{\vcenter{
    \offinterlineskip\ialign{\hbox to\dimexpr\@tempdima+1em{##}\cr
    \rightarrowfill\cr\noalign{\kern.5ex}
    \rightarrowfill\cr\noalign{\kern.5ex}
    \rightarrowfill\cr}}}\limits^{\!#1}}}
\makeatother

\newcommand{\beq}{\begin{equation}}
\newcommand{\eeq}{\end{equation}}

\makeatletter
\newcommand{\twoarrows}[3][0.2ex]{%
  \mathrel{\mathpalette\twoarrows@{{#1}{#2}{#3}}}%
}
\newcommand{\twoarrows@}[2]{\twoarrows@@#1#2}
\newcommand{\twoarrows@@}[4]{%
  \vcenter{\offinterlineskip\m@th
    \ialign{\hfil##\hfil\cr
      $#1#3$\cr
      \noalign{\vskip#2}
      $#1#4$\cr
    }%
  }%
}
\makeatother

\usepackage{algorithm}
\usepackage{algpseudocode}
\usepackage{graphicx}

\usepackage{multirow}
\usepackage{booktabs}
\usepackage{balance}
\usepackage{subfig}

\DeclareFontFamily{U}{dmjhira}{}
\DeclareFontShape{U}{dmjhira}{m}{n}{ <-> dmjhira }{}

\usepackage{natbib} 
\usepackage{mathtools} 
\usepackage{booktabs} 
\usepackage{tikz} 

\usepackage{tikz-cd}
\usetikzlibrary{arrows.meta,positioning}

\usepackage{booktabs}
\usepackage{tabularx}

\usepackage{listings}
\usepackage{xcolor}

\lstdefinelanguage{SQL}{
  morekeywords={SELECT,FROM,WHERE,JOIN,ON,GROUP,BY,ORDER,LIMIT,WITH,AS,COUNT,SUM,AVG,MIN,MAX,DISTINCT,CREATE,VIEW,INSERT,UPDATE,DELETE},
  sensitive=false,
  morecomment=[l]{--},
  morestring=[b]'
}

\usepackage{tcolorbox}

\usepackage{listings}
\usepackage{xcolor}
\title{\textsc{Csql}: Mapping Documents into Causal Databases
\thanks{Draft under revision.} }

\author{ Sridhar Mahadevan \\
	Adobe Research and University of Massachusetts, Amherst\\
	\texttt{smahadev@adobe.com, mahadeva@umass.edu}
}

\hypersetup{
pdftitle={A template for the arxiv style},
pdfsubject={q-bio.NC, q-bio.QM},
pdfauthor={David S.~Hippocampus, Elias D.~Striatum},
pdfkeywords={First keyword, Second keyword, More},
}

\begin{document}
\maketitle

\begin{abstract}
We describe a novel system, \textsc{Csql}, that automatically converts a collection of unstructured text documents into an SQL-queryable {\em causal database} (CDB). A CDB differs from a traditional DB: it is designed to answer ``why'' questions via causal interventions and structured causal queries. \textsc{Csql} builds on our earlier system, \textsc{Democritus}, which converts documents into thousands of local causal models derived from causal discourse \citep{mahadevan2025largecausalmodelslarge}. Unlike RAG-based systems or knowledge-graph--centric approaches, \textsc{Csql} supports causal analysis over document collections rather than purely associative retrieval. For example, given an article on the origins of human bipedal walking, \textsc{Csql} enables queries such as: ``What are the strongest causal influences on bipedalism?'' or ``Which variables act as causal hubs with the largest downstream influence?''

Beyond single-document case studies, we show that \textsc{Csql} can also ingest \emph{RAG/IE-compiled causal corpora} at scale by compiling the Testing Causal Claims (TCC) dataset of economics papers into a causal database containing 265{,}656 claim instances spanning 45{,}319 papers, 44 years, and 1{,}575 reported method strings, thereby enabling corpus-level causal queries and longitudinal analyses in SQL. Conceptually, \textsc{Csql} converts document collections into causally grounded relational databases, enabling causal analysis via standard SQL. In contrast to prior work that relies on hand-designed ontologies, fixed schemas, or domain-specific information extraction pipelines, \textsc{Csql} induces its schema directly from language (or from language-compiled causal artifacts). Viewed abstractly, \textsc{Csql} functions as a compiler from unstructured documents into a causal database equipped with a principled algebra of queries, and can be applied broadly across many domains ranging from business, economics, humanities, and science.
\end{abstract}


\section{Introduction}
\label{sec:intro} 

We introduce \textsc{Csql}, a causal knowledge discovery system that automatically converts collections of documents into SQL-queryable {\em causal} databases (CDB). Unlike retrieval-augmented generation (RAG) systems or knowledge-graph–centric approaches, which yield traditional databases, \textsc{Csql} supports \emph{causal analysis} over document collections rather than purely associative retrieval. \textsc{Csql} also differs fundamentally from traditional causal inference methods \citep{rubin-book,pearl-book}, which typically focus on narrow, domain-specific studies grounded in numerical or experimental data. In contrast, \textsc{Csql} operates directly on large collections of unstructured text.

A central feature of \textsc{Csql} is that it induces its relational schema directly from language. Unlike prior approaches that rely on hand-designed ontologies, fixed schemas, or domain-specific information extraction pipelines, \textsc{Csql} automatically constructs a causally grounded database whose structure emerges from discourse itself. Viewed abstractly, \textsc{Csql} functions as a compiler from unstructured documents into a causal database equipped with a principled algebra of queries. This design enables broad applicability across domains ranging from science and public policy to economics and history.

\begin{small} 
\begin{table*}[t]
\centering
\small
\begin{tabular}{lcccc}
\toprule
\textbf{Property} &
\textbf{RAG Systems} &
\textbf{Knowledge Graphs} &
\textbf{IE Pipelines} &
\textbf{\textsc{Csql}} \\
\midrule
Primary goal &
Answer generation &
Structured facts &
Entity / relation extraction &
Causal analysis \\

Input &
Text chunks &
Curated triples &
Annotated text &
Unstructured documents \\

Schema &
None (implicit) &
Predefined ontology &
Fixed schema &
\textbf{Induced from discourse} \\

Causality support &
None &
Implicit / informal &
Pairwise only &
\textbf{Weighted causal relations} \\

Uncertainty &
None &
Rare / ad hoc &
None &
\textbf{Aggregated model support} \\

Conflicting claims &
Not supported &
Manual resolution &
Not supported &
\textbf{Preserved and quantified} \\

Compositionality &
Prompt-based &
Limited graph traversal &
Limited &
\textbf{SQL joins and aggregations} \\

Counterfactuals &
No &
No &
No &
\textbf{Supported via causal structure} \\

Output format &
Generated text &
Graph database &
Triples / tables &
\textbf{Relational causal database} \\

Execution model &
Neural inference &
Symbolic querying &
Batch extraction &
\textbf{Deterministic SQL} \\
\bottomrule
\end{tabular}
\caption{Comparison of \textsc{Csql} with RAG systems, knowledge graphs, and traditional information extraction (IE) pipelines. \textsc{Csql} differs by compiling causal discourse into an SQL-queryable causal database with explicit uncertainty and support for causal analysis.}
\label{tab:comparison}
\end{table*}
\end{small}

Table~\ref{tab:comparison} contrasts traditional RAG or Knowledge Graph built databases with causal databases constructed by \textsc{Csql}.  \textsc{Csql} builds directly on our previous work on \textsc{Democritus}, which introduced an end-to-end system for discovering and evaluating causal structure from language. Given a document, \textsc{Democritus} automatically constructs thousands of local causal models, evaluates them for plausibility, and aggregates them into a ranked causal analysis expressed in natural language. \textsc{Csql} takes the output of \textsc{Democritus} one step further by compiling these causal artifacts into an SQL-queryable database, thereby enabling systematic causal querying over entire document collections.

\textsc{Csql} relies on recent advances in categorical causality \citep{mahadevan2025intuitionisticjdocalculustoposcausal,Fritz_2020} and geometric deep learning \citep{DBLP:conf/lics/FongST19,mahadevan2024gaiacategoricalfoundationsgenerative,gavranović2024positioncategoricaldeeplearning}, together with large language models (LLMs) used as discourse compilers. LLMs can generate rich causal narratives—enumerating subtopics, posing causal questions, and articulating mechanisms across domains ranging from macroeconomics to neuroscience. However, LLM-based document summarization alone cannot produce a causally structured, queryable database. \textsc{Csql} fills this gap by transforming causal discourse into an explicit relational representation.

For example, given a newspaper article—such as a recent \emph{Washington Post} story on whether eating dark chocolate may elongate life,\footnote{\url{https://www.washingtonpost.com/wellness/2025/12/26/dark-chocolate-health-benefits}}—\textsc{Csql} can construct a causal database that supports queries such as: ``What are the strongest causal influences on dark chocolate consumption?'' or ``Which variables act as downstream causal hubs in this discourse?'' More broadly, \textsc{Csql} is designed to support probing, comparative analysis of causal claims not only within individual documents, but across an entire \emph{frame of causal discourse} induced from a document collection.

A crucial design choice in \textsc{Csql} is the construction of this frame of discourse. Rather than restricting analysis to the input documents alone, the system treats each document as a seed for a broader causal neighborhood. \textsc{Democritus} first performs topic discovery over the input text and then expands these topics using a language model to construct a local discourse manifold. Causal claims are generated and evaluated relative to this expanded context. As a result, \textsc{Csql} reasons not only about what a document explicitly states, but about how its claims are situated within a larger landscape of causal discourse.

\section{The Origin of Bipedal Walking: A Running Example} 

\begin{figure}
    \centering
    \includegraphics[width=0.5\linewidth]{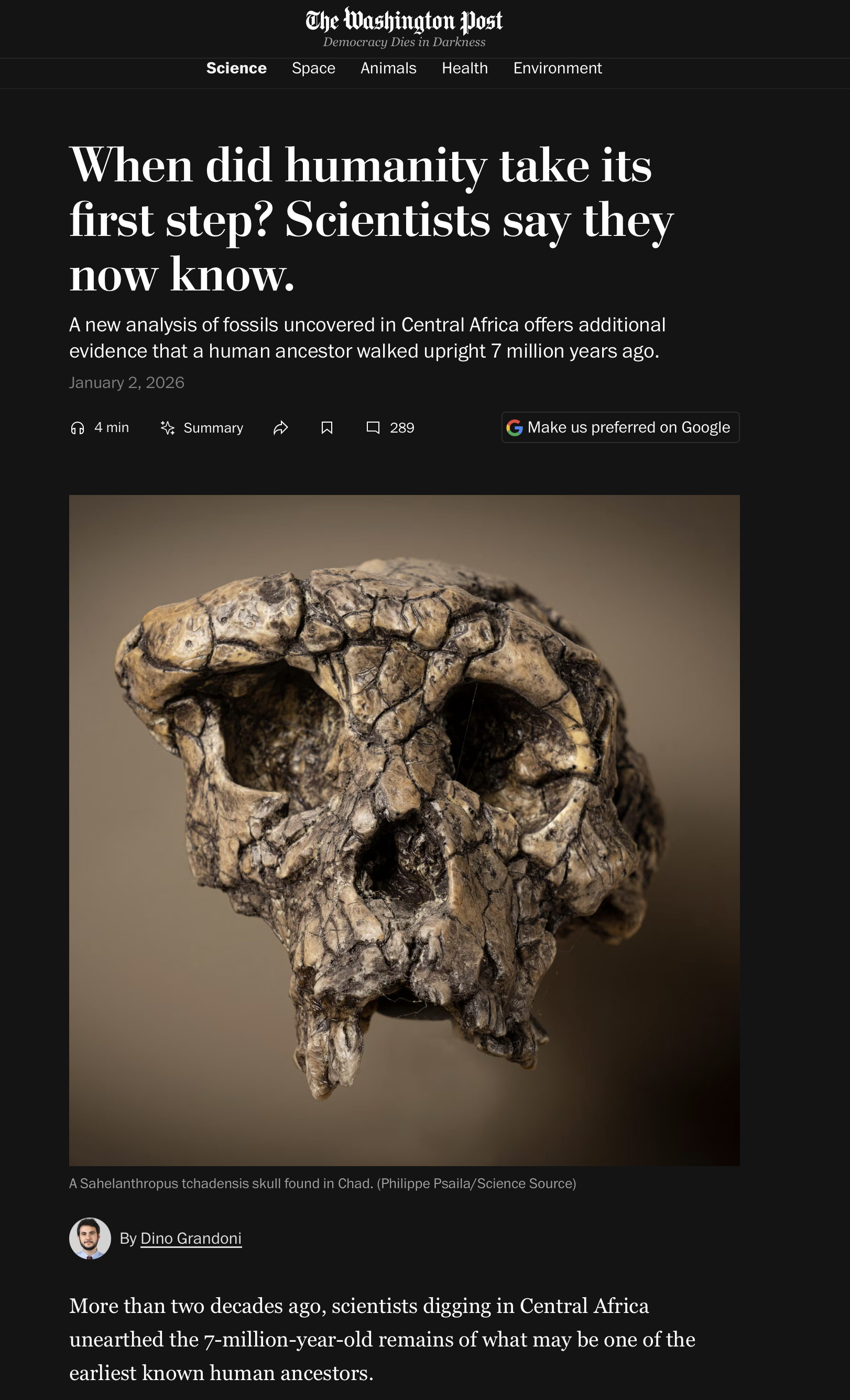}
    \caption{An article in The Washington Post on the origins of human bipedal walking.}
    \label{fig:WaPo_bipedal}
\end{figure}

\begin{figure}[p]
    \centering
    \includegraphics[width=0.75\linewidth]{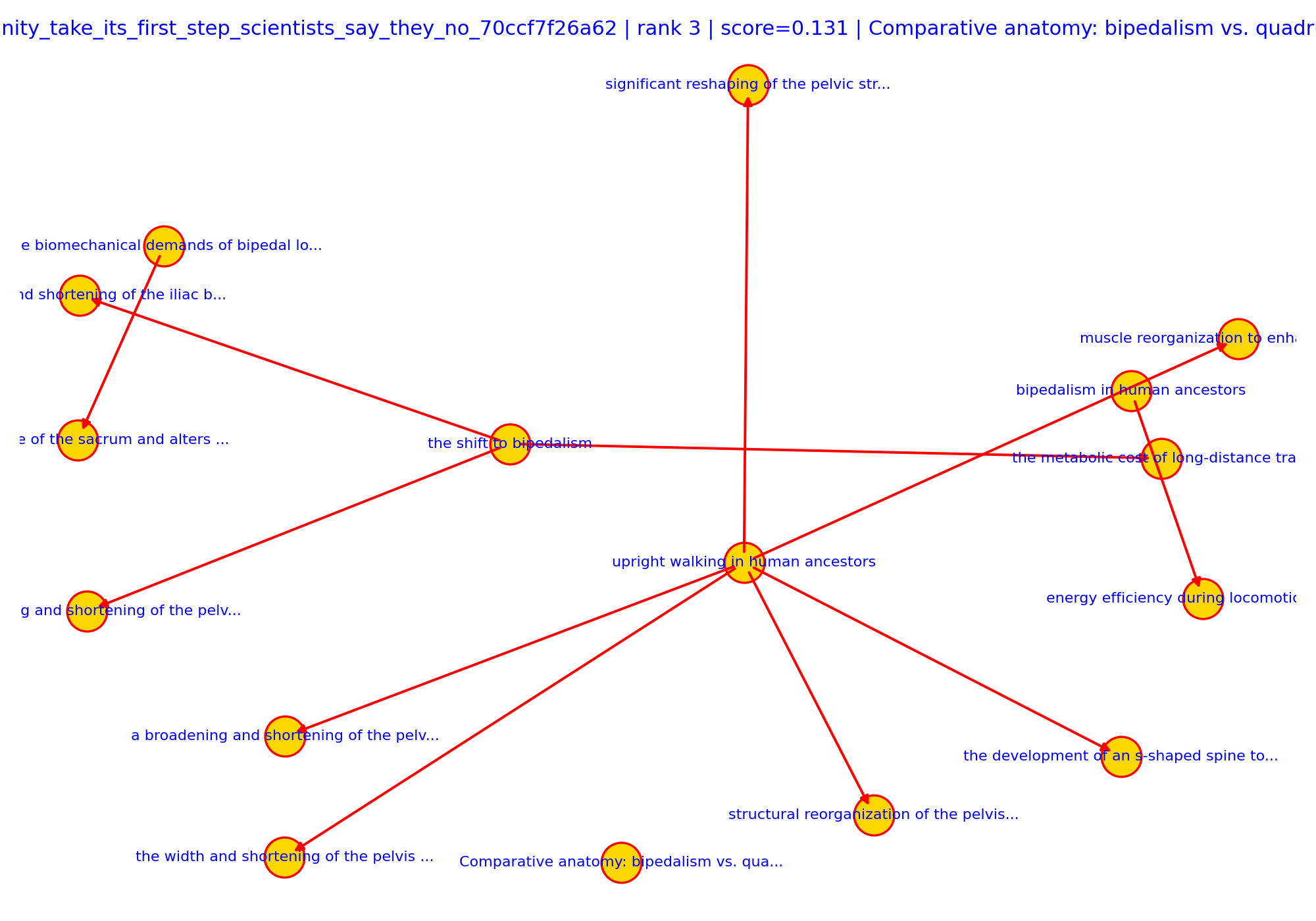}
    \includegraphics[width=0.75\linewidth]{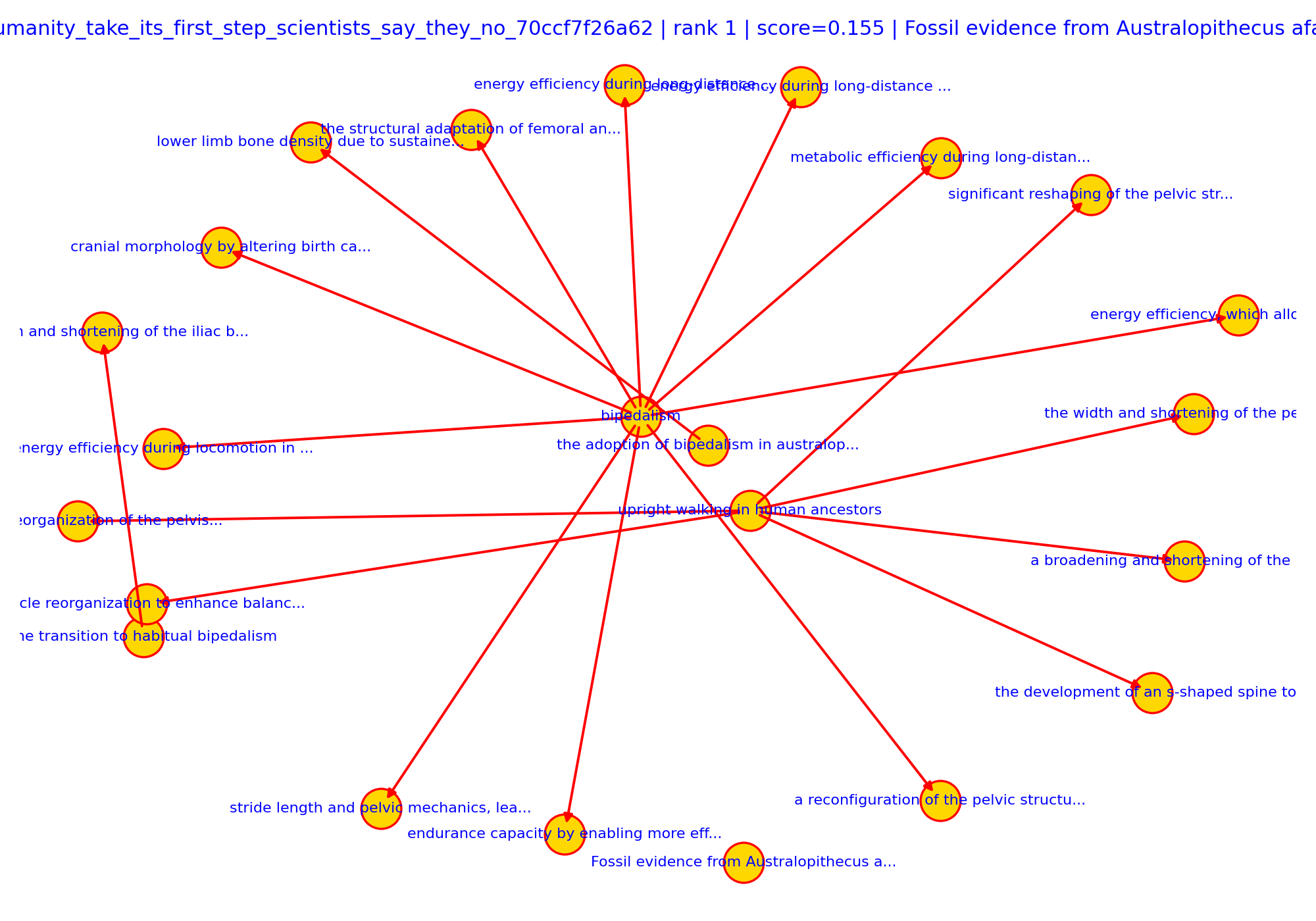}
    \caption{Two causal models constructed from The Washington Post article on the origins of bipedal walking by human ancestors 7 million years ago. \textsc{Democritus} constructs thousands of such models automatically from the PDF document, which are then converted by \textsc{Csql} into a causal database. }
    \label{fig:WaPo_LCMS}
\end{figure}

We will illustrate the construction of \textsc{Csql}  through a running example based on The Washington Post article on the origins of bipedal walking in human ancestors, dating back to $7$ million years ago (see Figure~\ref{fig:WaPo_bipedal}).\footnote{url{https://www.washingtonpost.com/science/2026/01/02/human-ancestor-biped/}}. The process of constructing a CSQL database involves first preprocessing the PDF document containing this article using our previous \textsc{Democritus} \citep{mahadevan2025largecausalmodelslarge} system.  \text{Democritus} is a system for building large causal DAG models  from carefully curated queries to a large language models (LLM) \citep{openai_chatgpt_release_notes,openai_gpt4o_system_card,anthropic_claude3_model_card,meta_llama3_herd,mistral_large2_2024}. \textsc{Democritus} can construct a rich library of DAG models from a text document in PDF format,  evaluate and rank order these claims quantitatively, and then produce a natural language executive summary. Example local causal models automatically constructed from this article are shown in Figure~\ref{fig:WaPo_LCMS}. 

\textsc{Democritus} constructs a detailed written summary of the causal claims both within the document, and the surrounding context that is automatically produced during the run. A brief except of such a report is given below. 

\begin{verbatim}
    # Causal Deep Dive —
    0001_when_did_humanity_take_its_first_step_scientists_say_they_no_70ccf7f26a62

This note is a human-readable interpretation of the Democritus credibility analysis.
It is **more detailed than an AI overview** but **less technical than the full model report**.

## What appears most credible

- upright walking in human ancestors —causes→ structural reorganization of the pelvis 
to support bipedal locomotion
- upright walking in human ancestors —increases→ the width and 
shortening of the pelvis to support bipedal locomotion
- upright walking in human ancestors —causes→ the development of 
an s-shaped spine to balance the body over the pelvis

\end{verbatim} 

In this paper, we instead convert the local causal models, such as shown in Figure~\ref{fig:WaPo_LCMS}, into a causal database over which SQL-like queries can be performed. We detail the CSQL data  model next.

\section{\textsc{Csql} Data Model}
\label{sec:data_model}

\subsection{Overview}

\textsc{Csql} represents causal knowledge extracted from document collections as a relational database.
Rather than storing isolated facts or entity relations, the database encodes \emph{causal structure}
together with empirical support aggregated across many local causal models. The data model is designed to satisfy three requirements:
(i) it must be queryable using standard SQL,
(ii) it must preserve uncertainty and disagreement present in the source documents,
and (iii) it must support compositional causal queries such as influence ranking,
hub detection, and multi-step causal paths.

\subsection{Core Relations}

A \textsc{Csql} database consists of the following core relations.

\paragraph{Nodes.}
Each distinct causal concept is represented as a node:

\begin{center}
\begin{tabular}{ll}
\texttt{node\_id} & 64-bit identifier \\
\texttt{label\_canon} & Canonicalized concept label \\
\texttt{label\_examples} & Example surface forms \\
\texttt{deg\_in} & Number of incoming causal edges \\
\texttt{deg\_out} & Number of outgoing causal edges
\end{tabular}
\end{center}

\paragraph{Edges.}
Causal relations are represented as weighted directed edges:

\begin{center}
\begin{tabular}{ll}
\texttt{edge\_id} & Unique edge identifier \\
\texttt{src\_id} & Source node \\
\texttt{dst\_id} & Destination node \\
\texttt{rel\_type} & Relation type (e.g., CAUSES, INFLUENCES) \\
\texttt{polarity} & Increase / decrease / unknown \\
\texttt{support\_lcms} & Number of supporting local models \\
\texttt{support\_docs} & Number of supporting documents \\
\texttt{score\_sum} & Aggregated credibility mass \\
\texttt{score\_mean} & Mean support per model \\
\texttt{score\_max} & Maximum single-model score
\end{tabular}
\end{center}

\paragraph{Edge Support.}

Fine-grained provenance is captured in a support table:

\begin{center}
\begin{tabular}{ll}
\texttt{edge\_id} & Referenced causal edge \\
\texttt{doc\_id} & Source document \\
\texttt{lcm\_instance\_id} & Supporting local model \\
\texttt{score\_raw} & Raw model score \\
\texttt{coupling} & Model-specific coupling strength
\end{tabular}
\end{center}

\subsection{Derived Relations}

\textsc{Csql} additionally materializes derived relations computed from the core tables.
These include:

\paragraph{Causal Modules (SCCs).}
Strongly connected components (SCCs) are computed over the causal graph:

\begin{center}
\begin{tabular}{ll}
\texttt{scc\_id} & Component identifier \\
\texttt{n\_nodes} & Number of nodes \\
\texttt{n\_edges} & Number of edges \\
\texttt{support\_docs} & Number of supporting documents \\
\texttt{top\_nodes} & High-degree nodes in the component
\end{tabular}
\end{center}

\subsection{Query Semantics}

\textsc{Csql} enables causal analysis through standard SQL queries.
For example:

\begin{itemize}
\item \textbf{Backbone extraction:} Identify the most credible causal relations by ordering edges by \texttt{score\_sum}.
\item \textbf{Causal hubs:} Identify nodes with large outgoing score mass.
\item \textbf{Downstream influence:} Join edges transitively to analyze multi-step causal paths.
\item \textbf{Disagreement analysis:} Compare \texttt{score\_max} and \texttt{score\_mean} to detect fragile claims.
\end{itemize}

All such queries are executed using deterministic SQL over materialized tables, rather than neural inference or prompt-based generation.

\section{Querying Causal Databases with \textsc{Csql}}
\label{sec:queries}

Once documents have been compiled into a \textsc{Csql} database, causal analysis is performed
using standard SQL queries.
Unlike neural or prompt-based systems, \textsc{Csql} queries execute deterministically over
materialized relations that encode causal structure, uncertainty, and provenance.

This section illustrates several representative query patterns supported by \textsc{Csql}.
All examples are executed using unmodified SQL over Parquet-backed tables and can be
run in any modern analytical database engine (e.g., DuckDB).

\subsection{Backbone Extraction}

A common analytic task is to identify the most credible causal relationships across
the document collection.
In \textsc{CSQL}, this corresponds to ordering causal edges by aggregated support.

\begin{lstlisting}[language=SQL]
SELECT
  e.rel_type,
  n1.label_canon AS src,
  n2.label_canon AS dst,
  e.support_lcms,
  e.score_sum
FROM atlas_edges e
JOIN atlas_nodes n1 ON e.src_id = n1.node_id
JOIN atlas_nodes n2 ON e.dst_id = n2.node_id
ORDER BY e.score_sum DESC
LIMIT 20;
\end{lstlisting} 

This query returns a \emph{causal backbone}: relationships that recur across many
high-scoring local causal models.
In the human bipedalism domain, this query surfaces relationships such as
\emph{bipedalism influencing metabolic efficiency} and
\emph{bipedalism increasing locomotion efficiency},
which are consistently supported across the extracted discourse.

\subsection{Causal Hubs}

\textsc{Csql}  supports identifying causal hubs: variables whose downstream influence is
large when aggregated across models.

\begin{lstlisting}[language=SQL]
SELECT
  n.label_canon AS src,
  SUM(e.score_sum) AS out_mass,
  COUNT(*) AS out_degree
FROM atlas_edges e
JOIN atlas_nodes n ON e.src_id = n.node_id
GROUP BY n.node_id, n.label_canon
ORDER BY out_mass DESC
LIMIT 10;
\end{lstlisting} 

This query ranks variables by the total credibility mass of their outgoing edges.
In practice, this identifies dominant explanatory drivers within a domain.
For example, in the bipedal walking corpus, \emph{bipedalism} emerges as a central
hub with broad downstream effects on energetics, skeletal structure, and locomotion.

\subsection{Local Mechanism Exploration}

\textsc{Csql} enables focused exploration of a single causal variable by filtering on a
specific source concept.

\begin{lstlisting}[language=SQL]
SELECT
  e.rel_type,
  n2.label_canon AS effect,
  e.support_lcms,
  e.score_sum
FROM atlas_edges e
JOIN atlas_nodes n1 ON e.src_id = n1.node_id
JOIN atlas_nodes n2 ON e.dst_id = n2.node_id
WHERE n1.label_canon = 'bipedalism'
ORDER BY e.score_sum DESC;
\end{lstlisting} 

This query enumerates the strongest downstream effects of a given cause,
ordered by credibility.
Such queries are useful for drilling into mechanistic hypotheses associated with
a particular concept, and can be composed with additional filters (e.g., relation
type or polarity).

\subsection{Provenance and Auditability}

Every causal edge in \textsc{Csql}  retains provenance information linking it to supporting
documents and local causal models.

\begin{lstlisting}[language=SQL]
SELECT
  s.doc_id,
  s.lcm_instance_id,
  s.score_raw,
  s.coupling
FROM atlas_edge_support s
WHERE s.edge_id = 10262796833405321330;
\end{lstlisting}

This query reveals exactly which documents and local models contributed to a given
causal claim.
Such provenance queries enable auditing, filtering by source, and sensitivity
analysis under alternative aggregation strategies.

\subsection{Cycles and Feedback Structures}

\textsc{Csql} supports detection of feedback and cyclic influence via derived relations such
as strongly connected components (SCCs).

\begin{lstlisting}[language=SQL]
SELECT
  scc_id,
  n_nodes,
  n_edges,
  support_docs,
  top_nodes
FROM atlas_scc
ORDER BY n_nodes DESC
LIMIT 5;
\end{lstlisting} 

Strongly connected components identify tightly coupled causal modules that may
represent feedback loops, co-evolving variables, or higher-level mechanisms.
These structures are difficult to express in DAG-based systems but arise naturally
in language-derived causal discourse.

\section{Example of a \textsc{Csql} Database}

\begin{figure*}[p]
\centering
\small
\setlength{\fboxsep}{6pt}

\begin{minipage}[t]{0.9\textwidth}
\textbf{(a) Backbone edges (top by score mass)}
\vspace{4pt}
\begin{tcolorbox}[colback=gray!3,colframe=black!35,boxrule=0.4pt]
\texttt{\footnotesize
SELECT e.rel\_type, n1.label\_canon AS src, n2.label\_canon AS dst, \\
\ \ \ \ \ \ e.support\_lcms, e.score\_sum\\
FROM read\_parquet('atlas\_edges.parquet') e\\
JOIN read\_parquet('atlas\_nodes.parquet') n1 ON e.src\_id=n1.node\_id\\
JOIN read\_parquet('atlas\_nodes.parquet') n2 ON e.dst\_id=n2.node\_id\\
ORDER BY e.score\_sum DESC\\
LIMIT 8;
}
\end{tcolorbox}

\vspace{-2pt}
\begin{tcolorbox}[colback=white,colframe=black!15,boxrule=0.4pt]
\texttt{\footnotesize
INFLUENCES | bipedalism | metabolic efficiency ... | 15 | 1.4624\\
INCREASES  | bipedalism | energy efficiency ...   | 13 | 1.3282\\
INCREASES  | bipedalism | energy efficiency ...   | 13 | 1.2644\\
INFLUENCES | bipedalism | endurance capacity ...  | 13 | 1.2557\\
INCREASES  | bipedalism | energy efficiency ...   | 12 | 1.2093\\
INFLUENCES | bipedalism | structural adaptation...| 11 | 1.2010\\
INFLUENCES | bipedalism | stride length ...       | 11 | 1.1029\\
INFLUENCES | reduced forest cover | selection ... |  6 | 0.4637\\
}
\end{tcolorbox}
\end{minipage}
\hfill

\begin{minipage}[t]{0.9\textwidth}
\textbf{(b) Causal hubs (sources with largest downstream mass)}
\vspace{4pt}
\begin{tcolorbox}[colback=gray!3,colframe=black!35,boxrule=0.4pt]
\texttt{\footnotesize
SELECT n.label\_canon AS src, SUM(e.score\_sum) AS out\_mass, COUNT(*) AS out\_deg\\
FROM read\_parquet('atlas\_edges.parquet') e\\
JOIN read\_parquet('atlas\_nodes.parquet') n ON e.src\_id=n.node\_id\\
GROUP BY n.node\_id, n.label\_canon\\
ORDER BY out\_mass DESC\\
LIMIT 8;
}
\end{tcolorbox}

\vspace{-2pt}
\begin{tcolorbox}[colback=white,colframe=black!15,boxrule=0.4pt]
\texttt{\footnotesize
bipedalism | 8.8239 | 10\\
reduced forest cover | 0.4637 | 1\\
changes in environmental conditions ... | 0.4026 | 1\\
abnormal femoral tubercle position | 0.2853 | 2\\
fossil structure of ardipithecus | 0.1358 | 1\\
muscle reorganization for balance ... | 0.1283 | 1\\
larger brain size in primates | 0.0814 | 1\\
lateral displacement of the femoral ... | 0.0763 | 1\\
}
\end{tcolorbox}
\end{minipage}
\hfill

\begin{minipage}[t]{0.9\textwidth}
\textbf{(c) Local mechanism: top outgoing edges from a hub}
\vspace{4pt}
\begin{tcolorbox}[colback=gray!3,colframe=black!35,boxrule=0.4pt]
\texttt{\footnotesize
SELECT e.rel\_type, n2.label\_canon AS dst, e.support\_lcms, e.score\_sum\\
FROM read\_parquet('atlas\_edges.parquet') e\\
JOIN read\_parquet('atlas\_nodes.parquet') n1 ON e.src\_id=n1.node\_id\\
JOIN read\_parquet('atlas\_nodes.parquet') n2 ON e.dst\_id=n2.node\_id\\
WHERE n1.label\_canon='bipedalism'\\
ORDER BY e.score\_sum DESC\\
LIMIT 7;
}
\end{tcolorbox}

\vspace{-2pt}
\begin{tcolorbox}[colback=white,colframe=black!15,boxrule=0.4pt]
\texttt{\footnotesize
INFLUENCES | metabolic efficiency ... | 15 | 1.4624\\
INCREASES  | energy efficiency ...    | 13 | 1.3282\\
INCREASES  | energy efficiency ...    | 13 | 1.2644\\
INFLUENCES | endurance capacity ...   | 13 | 1.2557\\
INCREASES  | energy efficiency ...    | 12 | 1.2093\\
INFLUENCES | structural adaptation... | 11 | 1.2010\\
INFLUENCES | stride length ...        | 11 | 1.1029\\
}
\end{tcolorbox}
\end{minipage}

\caption{
\textbf{\textsc{Csql} in action (DuckDB).} Example SQL queries over a Parquet-backed causal atlas
(\texttt{atlas\_nodes.parquet}, \texttt{atlas\_edges.parquet}). (a) Extracts backbone causal relations by aggregated
credibility mass. (b) Identifies causal hubs by total outgoing score mass. (c) Drills into a specific hub to recover its
highest-scoring downstream influences. All results are deterministic given the compiled atlas tables.
}
\label{fig:csql_queries}
\end{figure*}

Figure 3 illustrates how \textsc{Csql} represents causal knowledge extracted from a document collection as a relational database, using the example of upright walking in human ancestors. Each row in the \texttt{atlas\_edges} table corresponds to a causal relation discovered across multiple local causal models (LCMs), aggregated over the discourse frame induced by the source articles.

In this example, the concept \emph{bipedalism} emerges as a high-degree causal hub. The table shows that \emph{bipedalism} has strong outgoing causal influence on multiple downstream variables, including \emph{metabolic efficiency during long-distance locomotion}, \emph{energy efficiency during locomotion}, and \emph{endurance capacity}. These relations are not asserted from a single sentence or model; instead, each edge aggregates evidence across many independently generated LCMs, reflected in columns such as \texttt{support\_lcms} and \texttt{score\_sum}. Higher values indicate stronger redundancy and agreement across causal hypotheses.

Crucially, \textsc{Csql} exposes this causal structure through standard SQL. For example, a query selecting edges with maximal \texttt{score\_sum} directly answers questions such as: What are the strongest causal influences on bipedalism? Similarly, grouping by source nodes and summing outgoing score mass identifies causal hubs with large downstream impact. These operations require no specialized causal language or graph query engine—only relational joins and aggregations.

This example highlights the central design goal of \textsc{Csql}: to make causally grounded knowledge extracted from text accessible through familiar database abstractions. Rather than returning a flat list of extracted triples or a static knowledge graph, \textsc{Csql} materializes a causally meaningful schema in which credibility, redundancy, polarity, and structural roles (e.g., hubs, paths, cycles) can be queried, filtered, and composed using SQL. The following sections formalize this data model and show how increasingly sophisticated causal reasoning patterns arise naturally from relational queries over these tables.

These examples illustrate how \textsc{Csql} transforms document collections into
causally grounded relational databases.
By exposing causal structure, uncertainty, and provenance through standard SQL,
\textsc{Csql}  enables analysts to perform causal exploration, auditing, and hypothesis
generation using familiar database tools.

\subsection{Quantitative Properties of \textsc{Csql} Databases}
\label{sec:quantitative_csql}

Beyond supporting causal SQL queries, \textsc{Csql} induces a structured relational database whose
graph-theoretic and statistical properties can be summarized compactly.
These properties are useful for (i) diagnosing document-induced topic drift,
(ii) comparing domains, and (iii) choosing thresholds for downstream applications
(e.g., filtering low-support edges or selecting ``hub'' concepts for exploration).

\paragraph{Schema-level summary.}
For each document collection, \textsc{Csql} produces two primary relations:
(i) \texttt{atlas\_nodes} (canonicalized concept objects) and
(ii) \texttt{atlas\_edges} (canonicalized causal generators).
Edges carry both \emph{support} (how many local causal models contain the edge) and \emph{mass}
(e.g., \texttt{score\_sum}, \texttt{score\_mean}, \texttt{score\_max}) aggregated over model instances.
This provides a principled basis for ranking edges without imposing a fixed ontology.

\paragraph{Heavy-tailed support and score mass.}
Across domains we observe a consistently heavy-tailed distribution over edge support and score mass:
a small number of edges account for a disproportionate fraction of total score mass, while the
majority of edges have low support and low mass.
This is expected in discourse-derived causal graphs: a few mechanisms are stated (or paraphrased)
redundantly across discourse neighborhoods, while many peripheral claims occur only rarely.
In practice, this means aggressive compression is possible: for many analytic tasks,
retaining only the top $k$ edges by \texttt{score\_sum} preserves most of the interpretable backbone.

\paragraph{Hub concentration.}
\textsc{Csql}  makes it explicit that influence is concentrated around a small set of hubs.
Define the \emph{outgoing mass} of a node $x$ as
\[
M_{\mathrm{out}}(x) = \sum_{(x \xrightarrow{r} y)\in E} \mathrm{score\_sum}(x,r,y).
\]
In our bipedalism case study, the top hub (\texttt{bipedalism}) dominates outgoing mass by a wide margin.
We quantify this concentration using simple concentration ratios (e.g., top-1 and top-5 share of total
outgoing mass).

\paragraph{Relation-type mix and polarity mass.}
Edges are normalized into a small set of relation types (e.g., \texttt{CAUSES},
\texttt{INFLUENCES}, \texttt{INCREASES}, \texttt{REDUCES}) and polarity buckets
(\texttt{inc}, \texttt{dec}, \texttt{unk}).
We report the distribution of relation types and polarity mass as a coarse indicator of discourse framing.
For example, some domains emphasize \texttt{INFLUENCES}/\texttt{AFFECTS} (soft, explanatory discourse),
while others emphasize \texttt{CAUSES} or signed relations (\texttt{INCREASES}/\texttt{REDUCES}).

\paragraph{Cycles and strongly connected components.}
Unlike DAG-only representations, the induced causal database is not constrained to be acyclic.
We compute strongly connected components (SCCs) over the atlas edges to identify feedback loops and
closed influence motifs.
At the document scale, SCCs may be sparse; at corpus scale, SCC modules become a useful signal for
non-trivial causal connectivity and potential intervention loops.
We store SCC summaries in \texttt{atlas\_scc} when available.

\subsection{Counterfactual query via SQL view rewriting}

\textsc{Csql} supports intervention-style reasoning by treating interventions as deterministic rewrites of the atlas edge relation.
In the simplest \emph{hard intervention}, we apply a ``do-cut'' operator to a concept $X$ by deleting all outgoing causal generators with source $X$.
A counterfactual query is then evaluated as a \emph{difference of query results} between the baseline atlas and the intervened atlas view.
This provides a lightweight but concrete form of counterfactual reasoning: we can see which downstream influences disappear (or drop in rank) when the outgoing influence of a hub is removed. Figure~\ref{fig:csql_do_cut} shows the \textsc{Csql} specification for a counterfactual query. Table~\ref{tab:csql_do_cut_diff} shows the results of its application to the \textsc{Csql} database constructed for the running example of The Washington Post article on bipedal walking. 

\begin{figure*}[t]
\centering
\begin{minipage}{0.485\textwidth}
\lstset{language=SQL}
\begin{lstlisting}
-- Baseline: top outgoing generators from the hub "bipedalism"
SELECT
  e.rel_type,
  n2.label_canon AS dst,
  e.support_lcms,
  e.score_sum
FROM read_parquet('atlas_edges.parquet') e
JOIN read_parquet('atlas_nodes.parquet') n1 ON e.src_id = n1.node_id
JOIN read_parquet('atlas_nodes.parquet') n2 ON e.dst_id = n2.node_id
WHERE n1.label_canon = 'bipedalism'
ORDER BY e.score_sum DESC
LIMIT 10;
\end{lstlisting}
\end{minipage}
\hfill
\begin{minipage}{0.485\textwidth}
\lstset{language=SQL}
\begin{lstlisting}
-- Hard intervention (do-cut): delete all outgoing edges from "bipedalism"
WITH intervened_edges AS (
  SELECT e.*
  FROM read_parquet('atlas_edges.parquet') e
  JOIN read_parquet('atlas_nodes.parquet') n1 ON e.src_id = n1.node_id
  WHERE n1.label_canon <> 'bipedalism'
)
SELECT
  e.rel_type,
  n1.label_canon AS src,
  n2.label_canon AS dst,
  e.support_lcms,
  e.score_sum
FROM intervened_edges e
JOIN read_parquet('atlas_nodes.parquet') n1 ON e.src_id = n1.node_id
JOIN read_parquet('atlas_nodes.parquet') n2 ON e.dst_id = n2.node_id
ORDER BY e.score_sum DESC
LIMIT 10;
\end{lstlisting}
\end{minipage}

\vspace{0.5em}
\caption{\textbf{\textsc{Csql}  counterfactual reasoning as SQL query rewriting.}
Left: baseline query retrieves the strongest outgoing causal generators from the hub \texttt{bipedalism}.
Right: a hard intervention (\emph{do-cut}) is implemented as a view that removes all edges whose source is \texttt{bipedalism}, revealing the next-strongest hubs and mechanisms in the atlas.
All operations are deterministic transformations over the atlas tables.}
\label{fig:csql_do_cut}
\end{figure*}

\begin{table}[t]
\centering
\small
\begin{tabular}{llrr}
\toprule
\textbf{rel} & \textbf{dst (baseline top)} & \textbf{support} & \textbf{score\_sum}\\
\midrule
INFLUENCES & metabolic efficiency during long-distance travel  & 15 & 1.462 \\
INCREASES  & energy efficiency during locomotion in early hominins & 13 & 1.328 \\
INCREASES  & energy efficiency during long-distance locomotion     & 13 & 1.264 \\
INFLUENCES & endurance capacity (heat dissipation, etc.)           & 13 & 1.256 \\
INCREASES  & energy efficiency (metabolic resources)               & 12 & 1.209 \\
INFLUENCES & structural adaptation of femoral/tibial bones         & 11 & 1.201 \\
INFLUENCES & stride length and pelvic mechanics                    & 11 & 1.103 \\
\midrule
\multicolumn{4}{l}{\emph{Counterfactual (do-cut): these edges vanish from the atlas view.}} \\
\bottomrule
\end{tabular}
\caption{\textbf{Counterfactual diff for a hard intervention.}
Baseline: strongest outgoing generators from \texttt{bipedalism}.
Under do-cut(\texttt{bipedalism}) the outgoing edges are removed by construction, so these top-ranked consequences disappear from query results.}
\label{tab:csql_do_cut_diff}
\end{table}

\newpage 

It is also possible to implement ``soft" interventions in \textsc{Csql} as follows. 

\begin{lstlisting}[language=SQL] 
WITH soft_do AS (
  SELECT
    e.*,
    CASE WHEN n1.label_canon='bipedalism' THEN 0.2*e.score_sum ELSE e.score_sum END AS score_sum_do
  FROM read_parquet('atlas_edges.parquet') e
  JOIN read_parquet('atlas_nodes.parquet') n1 ON e.src_id=n1.node_id
)
SELECT ... ORDER BY score_sum_do DESC LIMIT 10;
\end{lstlisting}

\subsection{Quantitative Summary of a \textsc{Csql} Atlas}
\label{sec:quantitative}

\begin{table*}[t]
\centering
\small
\begin{tabular}{lcc}
\toprule
\multicolumn{3}{c}{\textbf{\textsc{Csql} Atlas Summary: WaPo Human Origins (Bipedalism)}} \\
\midrule
\multicolumn{3}{l}{\textbf{Atlas Size}} \\
\midrule
Number of nodes (concepts)          & \multicolumn{2}{c}{501} \\
Number of unique causal edges       & \multicolumn{2}{c}{65} \\
Edge-support rows (LCM evidence)    & \multicolumn{2}{c}{717} \\
\midrule
\multicolumn{3}{l}{\textbf{Hub Concentration}} \\
\midrule
Top hub                             & \multicolumn{2}{c}{\emph{bipedalism}} \\
Top hub outgoing mass share         & \multicolumn{2}{c}{0.748} \\
Top-5 hubs outgoing mass share      & \multicolumn{2}{c}{0.857} \\
\midrule
\multicolumn{3}{l}{\textbf{Edge Weight Distribution (score\_sum)}} \\
\midrule
Median ($p_{50}$)                   & \multicolumn{2}{c}{0.0337} \\
90th percentile ($p_{90}$)          & \multicolumn{2}{c}{0.847} \\
99th percentile ($p_{99}$)          & \multicolumn{2}{c}{1.377} \\
Tail ratio ($p_{99}/p_{50}$)        & \multicolumn{2}{c}{40.8} \\
\midrule
\multicolumn{3}{l}{\textbf{Relation-Type Mass Breakdown}} \\
\midrule
Relation type   & Polarity & Total score mass \\
\midrule
INFLUENCES      & unk      & 6.19 \\
INCREASES       & inc      & 5.13 \\
CAUSES          & unk      & 0.30 \\
REDUCES         & dec      & 0.14 \\
LEADS\_TO       & unk      & 0.03 \\
AFFECTS         & unk      & 0.01 \\
\bottomrule
\end{tabular}
\caption{
Quantitative summary of the \textsc{Csql} atlas induced from a Washington Post article on
the origins of human bipedal walking. The atlas exhibits strong hub concentration,
heavy-tailed edge weights, and a dominance of influence- and increase-type causal
relations. All statistics are computed directly via SQL queries over Parquet-backed
\textsc{Csql} tables using DuckDB.
}
\label{tab:csql_quantitative_summary}
\label{tab:WaPo_bipedal}
\end{table*}

To complement the qualitative examples, we report summary statistics of the
\textsc{Csql} atlas constructed from the Washington Post document on the origins of
bipedal walking. These statistics are computed directly over the atlas Parquet
tables using DuckDB SQL queries. The statistics are reported in Table~\ref{tab:WaPo_bipedal}. 

\paragraph{Atlas size.}
The atlas contains $501$ canonical concept nodes and $65$ unique canonical causal
edges. In addition, the edge-support table contains $717$ rows, where each row
records an instance of an atlas edge being supported by a particular local causal
model (LCM) in a particular document. This separation between (i) \emph{unique
canonical edges} and (ii) \emph{support instances} is essential: the atlas table
stores the distilled causal backbone, while the edge-support table retains
provenance for auditability and reproducibility.

\paragraph{Hub concentration.}
We measure hubness by aggregating outgoing causal mass per source concept
(\texttt{score\_sum} aggregated over outgoing edges). In this domain, the top hub
is \emph{bipedalism}. Remarkably, the top hub alone accounts for approximately
$74.8\%$ of the total outgoing score mass in the atlas, and the top five hubs
account for $85.7\%$. This demonstrates a strongly centralized causal atlas in
this document-induced discourse frame: most highly supported causal relations
radiate from a small number of discourse pivots.

\paragraph{Heavy-tailed edge strength.}
Edge strengths in \textsc{Csql} are strongly heavy-tailed. Using \texttt{score\_sum} as the
edge weight, the median (50th percentile) edge mass is $0.0337$, while the 90th
percentile is $0.847$, and the 99th percentile is $1.377$. The ratio
$p_{99}/p_{50} \approx 40.8$ quantifies the extreme concentration: a small number
of edges dominate the causal backbone, while most edges carry comparatively small
mass. This is consistent with the empirical behavior observed in Democritus
v2.0: causal discourse induces a large hypothesis space, but only a small
fraction of hypotheses are redundantly supported.

\paragraph{Relation-type and polarity mass.}
We also summarize atlas edges by normalized relation type (\texttt{rel\_type})
and polarity (\texttt{polarity}). In this atlas, most edges fall into two broad
families: \texttt{INFLUENCES} (unknown polarity) and \texttt{INCREASES} (positive
polarity). By mass, \texttt{INFLUENCES} contributes $6.19$ units of score mass,
while \texttt{INCREASES} contributes $5.13$, with smaller contributions from
\texttt{CAUSES}, \texttt{REDUCES}, and a handful of miscellaneous relations. This
breakdown reflects the linguistic character of the source document and the
implicit uncertainty of causal discourse: journalistic and paleoanthropological
claims are often phrased as influences and associations rather than signed,
quantified effects.

\paragraph{Top edges.}
Finally, we list the top weighted edges (by \texttt{score\_sum}). The highest-mass
edges take \emph{bipedalism} as the source and connect it to mechanistic
consequences such as metabolic efficiency, endurance capacity, stride length,
and locomotion energy efficiency. These edges collectively form the backbone of
the induced atlas.

\paragraph{Takeaway.}
These quantitative summaries illustrate why a \textsc{Csql}  atlas is useful: the induced
schema is not merely a flat list of extracted triples, but a weighted causal
database with (i) strong hub structure, (ii) heavy-tailed edge strength, and
(iii) an explicit decomposition by relation-type and polarity. Such structure is
immediately queryable by standard SQL, enabling downstream analytics (e.g.,
identifying dominant drivers, tracing multi-step influence chains, or locating
high-uncertainty edges) without requiring a hand-designed ontology or fixed
domain schema.

In summary, \textsc{Csql}  yields not only a queryable causal database but also a set of stable quantitative
signals (heavy tails, hub concentration, relation-type mixtures) that support corpus diagnostics and
downstream system design.

\subsection{Database Scale and Structure}

Table~\ref{tab:csql_scale} summarizes the size of the causal database
constructed from a single Washington Post article on the origins of
human bipedal walking.
Despite originating from a single document, the induced discourse
frame expands into hundreds of causal variables and hundreds of
supported causal relations.

\begin{table}[t]
\centering
\small
\begin{tabular}{lrr}
\toprule
\textbf{Quantity} & \textbf{Value} \\
\midrule
Number of nodes (concepts)        & 501 \\
Number of causal edges            & 65 \\
Number of edge-support records    & 717 \\
\bottomrule
\end{tabular}
\caption{Scale of the \textsc{Csql}  causal database constructed from a single
document.
Edge-support rows correspond to individual local causal models (LCMs)
that support each aggregated causal edge.}
\label{tab:csql_scale}
\end{table}

\subsection{Hub Dominance and Causal Centralization}

Causal influence in the atlas is highly concentrated around a small
number of hubs.
We quantify this concentration by measuring the fraction of total
outgoing causal mass attributable to the most influential source
nodes.

\begin{table}[t]
\centering
\small
\begin{tabular}{lcc}
\toprule
\textbf{Top hub} & \textbf{Top-1 mass share} & \textbf{Top-5 mass share} \\
\midrule
bipedalism & 0.748 & 0.857 \\
\bottomrule
\end{tabular}
\caption{Causal hub dominance in the \textsc{Csql} atlas.
The single most influential node (\texttt{bipedalism}) accounts for
nearly 75\% of outgoing causal mass, while the top five hubs account
for over 85\%.}
\label{tab:csql_hubs}
\end{table}

\subsection{Heavy-Tailed Causal Strength}

Causal edge strength in \textsc{Csql} follows a strongly heavy-tailed
distribution.
We characterize this using empirical quantiles of the aggregated
edge score distribution.

\begin{table}[t]
\centering
\small
\begin{tabular}{cccc}
\toprule
$p_{50}$ & $p_{90}$ & $p_{99}$ & $p_{99} / p_{50}$ \\
\midrule
0.034 & 0.847 & 1.377 & 40.8 \\
\bottomrule
\end{tabular}
\caption{Quantiles of causal edge score distribution.
The extreme $p_{99}/p_{50}$ ratio confirms a strongly heavy-tailed
energy landscape over causal hypotheses.}
\label{tab:csql_heavytail}
\end{table}

\subsection{Relation-Type Composition}

The causal database contains a heterogeneous mix of relation types,
with asymmetric distribution of causal mass across predicates.

\begin{table}[t]
\centering
\small
\begin{tabular}{lcr}
\toprule
\textbf{Relation type} & \textbf{\# edges} & \textbf{Total mass} \\
\midrule
INFLUENCES & 19 & 6.19 \\
INCREASES  & 27 & 5.13 \\
CAUSES     &  9 & 0.30 \\
REDUCES    &  7 & 0.14 \\
LEADS\_TO  &  1 & 0.03 \\
AFFECTS    &  2 & 0.01 \\
\bottomrule
\end{tabular}
\caption{Distribution of causal relation types.
Influence-style relations dominate both numerically and in total causal
mass, while strict causal predicates are rarer and more conservative.}
\label{tab:csql_reltypes}
\end{table}

\section{The \textsc{Csql} Data Model}
\label{sec:csql_data_model}

\textsc{Csql}  represents causally grounded knowledge extracted from document collections
as a small set of relational tables with well-defined semantics.
Unlike traditional information extraction pipelines, the schema is not
hand-designed in advance; instead, it is induced automatically from language
via the \textsc{Democritus} causal modeling pipeline.

At a high level, \textsc{Csql}  consists of four core relations:

\begin{itemize}
\item \texttt{atlas\_nodes}: canonical causal concepts,
\item \texttt{atlas\_edges}: aggregated causal relations,
\item \texttt{atlas\_edge\_support}: provenance linking edges to local causal models and documents,
\item \texttt{atlas\_scc}: strongly connected components capturing feedback structure.
\end{itemize}

Together, these tables define a causally meaningful relational schema that
supports both structural analysis and evidence-aware querying using standard SQL.

\subsection{Nodes: Canonical Causal Concepts}

Each row in \texttt{atlas\_nodes} represents a canonicalized causal concept
induced from language, such as \emph{bipedalism},
\emph{energy efficiency during locomotion}, or
\emph{metabolic efficiency during long-distance travel}.
Nodes are obtained by normalizing and de-duplicating subject and object phrases
across all extracted discourse triples.

Key columns include:
\begin{itemize}
\item \texttt{node\_id}: a stable identifier,
\item \texttt{label\_canon}: canonical concept string,
\item \texttt{label\_examples}: representative surface forms,
\item \texttt{deg\_in}, \texttt{deg\_out}: in- and out-degrees in the causal atlas.
\end{itemize}

Nodes function as the \emph{objects} of the induced causal structure.
No external ontology or controlled vocabulary is required; the object space is
learned directly from text.

\subsection{Edges: Aggregated Causal Relations}

Each row in \texttt{atlas\_edges} represents a causal relation aggregated across
many local causal models (LCMs). An edge corresponds to a canonical triple
\[
(\texttt{src\_id}, \texttt{rel\_type}, \texttt{dst\_id}),
\]
where \texttt{rel\_type} ranges over normalized relations such as
\texttt{INCREASES}, \texttt{REDUCES}, \texttt{CAUSES}, or \texttt{INFLUENCES}.

In addition to structural information, each edge carries quantitative evidence:
\begin{itemize}
\item \texttt{support\_lcms}: number of LCMs in which the edge appears,
\item \texttt{support\_docs}: number of documents contributing evidence,
\item \texttt{score\_sum}, \texttt{score\_mean}, \texttt{score\_max}: aggregated credibility scores,
\item polarity mass fields distinguishing increasing, decreasing, and unknown effects.
\end{itemize}

These values encode \emph{agreement under variation}: edges that recur across
many high-scoring hypotheses accumulate greater score mass and therefore higher
causal credibility.

\subsection{Edge Support and Provenance}

\textsc{Csql} preserves fine-grained provenance via the \texttt{atlas\_edge\_support} table,
which links each aggregated edge back to:
\begin{itemize}
\item specific LCM instances,
\item source documents,
\item raw per-model scores and coupling values.
\end{itemize}

This design enables auditability and traceability.
Users can drill down from a high-level causal claim to the specific models and
textual evidence that support it.

\subsection{Strongly Connected Components}

The \texttt{atlas\_scc} table captures strongly connected components of the causal
atlas graph. These components represent feedback loops or tightly coupled causal
subsystems, which cannot be represented in DAG-based causal formalisms.

SCCs provide a natural entry point for reasoning about cyclic causality, mutual
reinforcement, and dynamical regimes—key motivations for moving beyond DAGs to
more expressive causal representations.

\section{Causal Reasoning as SQL}
\label{sec:csql_queries}

A central contribution of \textsc{Csql}  is that \emph{causal reasoning can be expressed
directly as SQL queries} over the induced schema.
No specialized graph query language or causal DSL is required.

\subsection{Identifying Causal Backbones}

The strongest causal claims in a document collection correspond to edges with
maximal aggregated score mass. For example:

\begin{lstlisting}[language=SQL]
SELECT
  e.rel_type,
  n1.label_canon AS src,
  n2.label_canon AS dst,
  e.support_lcms,
  e.score_sum
FROM atlas_edges e
JOIN atlas_nodes n1 ON e.src_id = n1.node_id
JOIN atlas_nodes n2 ON e.dst_id = n2.node_id
ORDER BY e.score_sum DESC
LIMIT 50;
\end{lstlisting}

This query extracts the \emph{causal backbone} of the discourse frame—relations
that survive across many high-scoring hypotheses.

\subsection{Causal Hubs and Downstream Influence}

Causal hubs are nodes with large outgoing score mass:

\begin{lstlisting}[language=SQL]
SELECT
  n.label_canon AS src,
  SUM(e.score_sum) AS out_mass
FROM atlas_edges e
JOIN atlas_nodes n ON e.src_id = n.node_id
GROUP BY n.label_canon
ORDER BY out_mass DESC;
\end{lstlisting}

In the upright walking example, \emph{bipedalism} emerges as the dominant hub,
with strong downstream influence on metabolic efficiency, endurance, and
locomotion.

\subsection{Causal Composition}

Multi-step causal chains correspond to relational joins:

\begin{lstlisting}[language=SQL]
SELECT
  n1.label_canon AS a,
  e1.rel_type AS r1,
  n2.label_canon AS b,
  e2.rel_type AS r2,
  n3.label_canon AS c,
  (e1.score_sum + e2.score_sum) AS path_score
FROM atlas_edges e1
JOIN atlas_edges e2 ON e1.dst_id = e2.src_id
JOIN atlas_nodes n1 ON e1.src_id = n1.node_id
JOIN atlas_nodes n2 ON e1.dst_id = n2.node_id
JOIN atlas_nodes n3 ON e2.dst_id = n3.node_id
ORDER BY path_score DESC;
\end{lstlisting}

Such queries implement causal composition directly in SQL.

\subsection{Cycles and Feedback}

Cycles are detected via self-joins or SCC tables. For example, mutual influence:

\begin{lstlisting}[language=SQL]
SELECT
  n1.label_canon AS a,
  e1.rel_type AS r1,
  n2.label_canon AS b,
  e2.rel_type AS r2
FROM atlas_edges e1
JOIN atlas_edges e2
  ON e1.src_id = e2.dst_id
 AND e1.dst_id = e2.src_id
JOIN atlas_nodes n1 ON e1.src_id = n1.node_id
JOIN atlas_nodes n2 ON e1.dst_id = n2.node_id;
\end{lstlisting}

This enables explicit querying of feedback structures absent from
DAG-based causal systems.

\paragraph{Counterfactual query via SQL view rewriting.}
\textsc{Csql}  supports intervention-style reasoning by treating interventions as deterministic rewrites of the atlas edge relation.
In the simplest \emph{hard intervention}, we apply a ``do-cut'' operator to a concept $X$ by deleting all outgoing causal generators with source $X$.
A counterfactual query is then evaluated as a \emph{difference of query results} between the baseline atlas and the intervened atlas view.
This provides a lightweight but concrete form of counterfactual reasoning: we can see which downstream influences disappear (or drop in rank) when the outgoing influence of a hub is removed.

\subsection{Quantitative Summary of the \textsc{Csql}  Corpus Database}
\label{sec:quant-csql}

Table~\ref{tab:csql_corpus_summary} reports corpus-level summary statistics for the
merged \textsc{Csql} database constructed from multiple document-derived atlases. The database
contains canonicalized concept nodes and relation edges, together with provenance
(edge-support rows linking edges to document and local-model instances) and aggregate
credibility statistics (e.g., score mass and polarity mass). The resulting score
distribution is heavy-tailed, and the outgoing score mass is concentrated in a
small number of causal hubs, reflecting the system’s tendency to surface central
drivers in the discourse manifold.

\begin{table}[t]
\centering
\small
\begin{tabular}{l r}
\toprule
\textbf{Quantity} & \textbf{Value} \\
\midrule
Canonical nodes ($|\mathrm{nodes}|$) & 839 \\
Canonical edges ($|\mathrm{edges}|$) & 450 \\
Edge-support rows ($|\mathrm{edge\_support}|$) & 1311 \\
Documents ($|\mathrm{docs}|$) & 10 \\
Atlases merged ($|\mathrm{atlases}|$) & 7 \\
\bottomrule
\end{tabular}
\caption{Corpus-level size statistics for the merged \textsc{Csql} database. Nodes and edges
are canonicalized (string-normalized) concepts and relations; edge-support rows store
provenance at the level of (document, local causal model) instances.}
\label{tab:csql_corpus_summary}
\end{table}

\begin{table}[t]
\centering
\small
\begin{tabular}{l r r r}
\toprule
\textbf{Hub source (src)} & \textbf{Out mass} & \textbf{Out edges} & \textbf{LCM support} \\
\midrule
exposure to roundup & 105.03 & 3 & 24 \\
dark chocolate consumption & 64.70 & 5 & 20 \\
severity of post impacts & 44.76 & 2 & 17 \\
impact of researchers & 33.68 & 3 & 15 \\
application of roundup & 30.58 & 2 & 9 \\
\bottomrule
\end{tabular}
\caption{Top causal hubs by outgoing score mass in the merged \textsc{Csql} corpus database.}
\label{tab:csql_hubs2}
\end{table}

\begin{table}[t]
\centering
\small
\begin{tabular}{l l r r}
\toprule
\textbf{rel\_type} & \textbf{polarity} & \textbf{\#edges} & \textbf{score mass} \\
\midrule
INFLUENCES & unk & 135 & 353.78 \\
LEADS\_TO  & unk & 80  & 188.11 \\
REDUCES    & dec & 75  & 181.83 \\
INCREASES  & inc & 89  & 156.30 \\
CAUSES     & unk & 50  & 89.99 \\
AFFECTS    & unk & 21  & 75.71 \\
\bottomrule
\end{tabular}
\caption{Distribution of edge mass by normalized relation type and polarity in the merged corpus.}
\label{tab:csql_rel_mass}
\end{table}

\paragraph{Interpretation.}
The merged \textsc{Csql}  database exhibits two recurring properties. First, the aggregate
credibility signal (score mass) is heavy-tailed: a small fraction of edges and
sources account for a large fraction of total mass. Second, this mass is organized
around a handful of causal hubs (Table~\ref{tab:csql_hubs}), which act as central
drivers in the discourse manifold (e.g., \emph{exposure to roundup} and \emph{dark
chocolate consumption}). These hubs are precisely the nodes that become most useful
for interactive causal querying: they support SQL queries that surface high-impact
downstream consequences, compare competing mechanisms, and identify which claims are
stable across multiple local causal models. Finally, the relation-mass breakdown
(Table~\ref{tab:csql_rel_mass}) shows that the induced schema concentrates into a
small, reusable vocabulary of causal predicates (INFLUENCES/LEADS\_TO/INCREASES/REDUCES),
which is critical for portability across corpora without requiring a hand-built ontology.

\section{Algorithmic Construction of \textsc{Csql} Databases}
\label{sec:csql_algorithms}

This section describes the deterministic compilation step that turns a collection
of local causal models (LCMs) into a \emph{causal SQL database} (\textsc{Csql}). The input to this
stage is a directory of LCMs produced by an upstream system (e.g., \textsc{Democritus})
and optionally accompanied by scoring metadata (e.g., \texttt{scores.csv}). The output is a small
set of Parquet tables that can be queried directly in DuckDB (or any Parquet-capable
SQL engine).

\subsection{Input Contract}
\label{sec:csql_input}

We assume each document $d$ has an associated set of LCMs stored as JSON files.
An LCM is represented as a directed labeled multigraph
$M = (V_M, E_M)$ with a focus string and optional metadata. Each edge is a record
\[
e = (\texttt{src}, \texttt{rel}, \texttt{dst})
\]
where \texttt{src} and \texttt{dst} are surface strings and \texttt{rel} is a free-form relation phrase
(e.g., ``increases'', ``reduces'', ``influences''). Optionally, each LCM is accompanied by a scalar score
and auxiliary fields (radius, model size, coupling, etc.) used for aggregation.

\subsection{\textsc{Csql} Schema}
\label{sec:csql_schema}

\textsc{Csql} materializes three core tables (plus one optional analytic table):

\begin{itemize}
    \item \textbf{\texttt{nodes}}: canonicalized concepts (entities/variables) discovered from LCM node strings.
    \item \textbf{\texttt{edges}}: canonicalized causal edges aggregated across all LCMs (deduplicated).
    \item \textbf{\texttt{edge\_support}}: provenance table linking each canonical edge to the document and LCM instances
    that support it (including local model scores).
    \item \textbf{\texttt{scc}} (optional): strongly connected component summaries over the induced graph, to expose cycles/modules.
\end{itemize}

This separation is deliberate and mirrors standard database normalization practice:
\texttt{edges} provides a stable fact table for queries, while \texttt{edge\_support} preserves provenance and enables
auditing, filtering, and corpus merging.

\subsection{Canonicalization and Keying}
\label{sec:canonicalization}

To make the database queryable and mergeable across documents, \textsc{Csql} normalizes strings and relations.
We define:

\paragraph{Canonical node labels.}
A normalization function $\mathrm{canon}(\cdot)$ maps raw node strings to a canonical form
(lowercasing, whitespace/punctuation normalization, dash unification, etc.). Canonical nodes are keyed by:
\[
\texttt{node\_id} = H(\mathrm{canon}(x))
\]
where $H$ is a stable 64-bit hash.

\paragraph{Canonical relation types.}
A relation normalization function $\mathrm{reltype}(\cdot)$ maps raw relation phrases to a small controlled vocabulary
(e.g., \textsc{CAUSES}, \textsc{INFLUENCES}, \textsc{INCREASES}, \textsc{REDUCES}, \textsc{AFFECTS}, \textsc{LEADS\_TO}).
We also extract a coarse \emph{polarity} label $\in \{\texttt{inc},\texttt{dec},\texttt{unk}\}$.

\paragraph{Canonical edges.}
A canonical edge is identified by:
\[
\texttt{edge\_id} = H(\texttt{src\_id} \,\|\, \texttt{rel\_type} \,\|\, \texttt{dst\_id})
\]
This provides stable deduplication within and across documents.

\subsection{Atlas Builder: From LCMs to Parquet}
\label{sec:atlas_builder}

Algorithm~\ref{alg:build_atlas} describes the \textsc{Csql} compilation procedure. The output tables store:
(i) canonical objects and arrows (\texttt{nodes}, \texttt{edges}),
(ii) provenance and scoring (\texttt{edge\_support}),
and (iii) optional graph analytics (\texttt{scc}).

\begin{algorithm}[t]
\caption{\textsc{BuildAtlas}: Compile LCMs into a \textsc{Csql} database}
\label{alg:build_atlas}
\begin{algorithmic}[1]
\Require
Document runs root directory $\mathcal{R}$ containing per-document LCM folders;
optional scoring metadata (\texttt{scores.csv}); min-edge filter $\tau$; relation whitelist/blacklist.
\Ensure
Parquet tables: \texttt{nodes.parquet}, \texttt{edges.parquet}, \texttt{edge\_support.parquet} (and optional \texttt{scc.parquet}).

\State Initialize empty maps: $\texttt{NodeMap}: \texttt{node\_id}\mapsto$ node record, $\texttt{EdgeAgg}: \texttt{edge\_id}\mapsto$ aggregate record
\State Initialize empty list $\texttt{SupportRows}$

\ForAll{documents $d$ under $\mathcal{R}$}
    \State Load set of LCM JSON files $\{M_i\}$ for $d$ (optionally filtered by score/size)
    \ForAll{LCMs $M_i$}
        \State Extract metadata: $\texttt{doc\_id}$, $\texttt{lcm\_instance\_id}$, $\texttt{score}$, $\texttt{score\_raw}$, $\texttt{coupling}$
        \ForAll{edges $e=(u,r,v)\in E_{M_i}$}
            \If{$|E_{M_i}| < \tau$} \textbf{continue} \EndIf
            \State $u_c \gets \mathrm{canon}(u)$; $v_c \gets \mathrm{canon}(v)$
            \State $\texttt{src\_id}\gets H(u_c)$; $\texttt{dst\_id}\gets H(v_c)$
            \State $(\texttt{rel\_type},\texttt{polarity}) \gets \mathrm{reltype}(r)$
            \State $\texttt{edge\_id}\gets H(\texttt{src\_id}\,\|\,\texttt{rel\_type}\,\|\,\texttt{dst\_id})$
            \State Update $\texttt{NodeMap}$ with examples for $u, v$
            \State Update $\texttt{EdgeAgg}[\texttt{edge\_id}]$:
                support counts, score mass sums, polarity mass
            \State Append support row to $\texttt{SupportRows}$:\\
            \begin{eqnarray*}
(\texttt{edge\_id},\texttt{doc\_id},\texttt{atlas\_id}, \texttt{lcm\_instance\_id},\texttt{score},\texttt{score\_raw},\texttt{coupling}) 
     \end{eqnarray*}
        \EndFor
    \EndFor
\EndFor

\State Materialize $\texttt{nodes}$ table from $\texttt{NodeMap}$ with degree stats computed from $\texttt{edges}$
\State Materialize $\texttt{edges}$ table from $\texttt{EdgeAgg}$ with summary fields:
\Statex \quad 
\begin{eqnarray*} \texttt{support\_lcms},\texttt{support\_docs},\texttt{score\_sum},\texttt{score\_mean}, \\ \texttt{score\_max},\texttt{pol\_mass\_inc},\texttt{pol\_mass\_dec},\texttt{pol\_mass\_unk},\texttt{controversy} \end{eqnarray*}
\State Materialize $\texttt{edge\_support}$ table from $\texttt{SupportRows}$

\State (Optional) Build induced directed graph from \texttt{edges}; compute SCCs; write \texttt{scc.parquet}
\State Write all tables to Parquet
\State \Return
\end{algorithmic}
\end{algorithm}

\paragraph{Aggregation and ``controversy''.}
Each canonical edge aggregates evidence across many LCM instances. Let $w_i$ be the score mass contributed by LCM $i$
to a canonical edge (we use the adjusted score when available). For polarity mass we define:
\[
m_{\texttt{inc}} = \sum_{i:\,\texttt{polarity}=\texttt{inc}} w_i,\quad
m_{\texttt{dec}} = \sum_{i:\,\texttt{polarity}=\texttt{dec}} w_i,\quad
m_{\texttt{unk}} = \sum_{i:\,\texttt{polarity}=\texttt{unk}} w_i.
\]
A simple controversy indicator is then:
\[
\mathrm{controversy} = \frac{\min(m_{\texttt{inc}}, m_{\texttt{dec}})}{m_{\texttt{inc}} + m_{\texttt{dec}} + \epsilon}.
\]
This yields $\mathrm{controversy}=0$ when only one direction is supported and approaches $0.5$ when inc/dec are balanced.

\subsection{Corpus Merge: Union of \textsc{Csql} Databases}
\label{sec:csql_merge}

Because all identifiers are canonical (hash-based), merging multiple atlas databases is a deterministic union
operation followed by re-aggregation. Algorithm~\ref{alg:merge_atlas} sketches the merge procedure.

\begin{algorithm}[t]
\caption{\textsc{MergeAtlases}: Merge per-document \textsc{Csql} databases into a corpus database}
\label{alg:merge_atlas}
\begin{algorithmic}[1]
\Require A directory containing atlas folders $\{\mathcal{A}_j\}$, each with \texttt{nodes}, \texttt{edges}, \texttt{edge\_support}.
\Ensure Merged corpus \textsc{Csql} database: \texttt{nodes}, \texttt{edges}, \texttt{edge\_support}.

\State Load and concatenate all \texttt{edge\_support} tables; optionally prefix $\texttt{doc\_id}$ with atlas id for disambiguation
\State Group \texttt{edge\_support} by \texttt{edge\_id} to recompute edge aggregates (support counts, score sums, polarity mass)
\State Construct merged \texttt{edges} from aggregated groups
\State Construct merged \texttt{nodes} by collecting all referenced \texttt{src\_id}, \texttt{dst\_id} and recomputing degrees from merged \texttt{edges}
\State Write merged tables to Parquet
\State \Return
\end{algorithmic}
\end{algorithm}

In this way, \textsc{Csql}  supports both \emph{document-level} analysis and \emph{corpus-level} analysis using the same schema.
Crucially, provenance is preserved: corpus queries can be refined by filtering on \texttt{doc\_id}, \texttt{atlas\_id}, or \texttt{lcm\_instance\_id},
enabling traceable drill-down from aggregate causal claims back to local models and source documents.

\subsection{Practical Notes for Users}
\label{sec:csql_practical}

\textsc{Csql} is designed to be usable with off-the-shelf tools:
\begin{itemize}
    \item Any SQL engine with Parquet support can query the tables; we use DuckDB.
    \item Most queries are expressed as joins between \texttt{edges} and \texttt{nodes} for human-readable labels.
    \item Provenance-aware queries join \texttt{edge\_support} to recover document and LCM-level evidence.
\end{itemize}
This makes \textsc{Csql} a \emph{causal database backend} that can be placed under visualization layers, RAG systems,
or higher-level reasoning systems (including \textsc{Democritus} itself).

\section{\textsc{Csql} from RAG-Compiled Causal Corpora}
\label{sec:csql_from_rag}

A key design goal of \textsc{Csql} is to decouple \emph{causal discourse compilation}
from \emph{causal database construction}. While earlier sections focused on
LLM-driven discourse compilation via \textsc{Democritus}, the CSQL pipeline is
agnostic to how causal claims are obtained. In this subsection, we show that
\textsc{CSQL} can be constructed directly from \emph{RAG-compiled causal corpora},
without requiring access to the original documents or an LLM-based generation
pipeline.

We use the \emph{Testing Causal Claims (TCC)} dataset \citep{garg_fetzer_2025} as a
canonical example. TCC extracts causal claims from a large corpus of economics
papers using information extraction and retrieval-based methods, yielding
structured records of the form
\[
(\text{cause}, \text{effect}, \text{sign}, \text{method}, \text{document metadata}).
\]
Rather than treating TCC as a static knowledge graph, we interpret it as a
\emph{precompiled causal discourse layer}. \textsc{Csql} then compiles this layer
into a relational causal database supporting causal aggregation, ranking, and
intervention-style queries.

\paragraph{Discourse-to-database compilation.}
Given a RAG-compiled corpus such as TCC, \textsc{Csql} performs a deterministic
compilation step that maps extracted causal claims into a small set of normalized
relational tables. Canonicalized cause and effect phrases become nodes, while
typed causal relations become edges. Repeated claims across documents are
aggregated, producing support counts and score mass analogous to those obtained
from \textsc{Democritus}-generated local causal models.

Importantly, \textsc{Csql} does not assume a fixed ontology, predefined schema, or
hand-designed knowledge graph. Instead, the schema is induced directly from the
causal discourse itself. Provenance information—such as document identifiers,
publication year, and causal inference method—is preserved in an auxiliary
edge-support table, enabling downstream filtering and auditability.

\paragraph{Unified causal querying across compilation methods.}
Once compiled, a \textsc{Csql} database constructed from a RAG-based corpus supports the
same class of causal queries as one constructed from \textsc{Democritus}. For
example:
\begin{itemize}
    \item identifying global causal hubs with the largest downstream influence,
    \item ranking causal claims by redundancy and aggregate support,
    \item detecting controversial claims with mixed directional evidence,
    \item approximating interventions by removing or conditioning on selected
    sources.
\end{itemize}
From the perspective of the query layer, there is no distinction between causal
databases derived from LLM-generated discourse and those derived from
RAG-compiled corpora.

\paragraph{Implications for scale and interoperability.}
This separation enables \textsc{Csql} to operate at corpus scale without requiring
LLM inference over tens of thousands of documents. Large, curated resources such
as TCC can be ingested directly, allowing \textsc{Csql} to support causal analysis
over orders of magnitude more documents than would be feasible with generative
pipelines alone. At the same time, databases compiled from RAG-based sources and
from \textsc{Democritus} can be merged seamlessly, yielding a unified causal
database spanning heterogeneous document collections.

Viewed abstractly, \textsc{Csql} serves as a common compilation target for diverse
causal discourse generators—LLMs, RAG systems, and classical information
extraction pipelines—while providing a uniform, SQL-native interface for causal
analysis.

\subsection{\textsc{Csql} from a RAG-Compiled Causal Corpus: Testing Causal Claims (TCC)}
\label{sec:tcc_csql_results}

We also instantiate \textsc{Csql} on the \emph{Testing Causal Claims} (TCC) dataset, which already provides a
corpus-level extraction of directed cause--effect claims from $\sim$45K economics papers.
The \emph{Testing Causal Claims} (TCC) project \citep{garg_fetzer_2025} extracts and
catalogs causal claims from large corpora of economics papers, producing a global
index of cause--effect statements enriched with metadata such as sign, statistical
significance, and inference method.   In this setting, \textsc{Csql} does not run the \textsc{Democritus} discourse compiler; instead, it treats TCC as a
pre-compiled discourse layer and compiles it into the same relational schema
(\texttt{nodes}, \texttt{edges}, \texttt{edge\_support}).

The resulting database contains 295{,}459 canonicalized concept strings and 260{,}777 distinct directed
edges (265{,}656 edge-support rows), spanning 44 publication years. Because the current ingest uses the
generic relation label \texttt{INFLUENCES} and does not yet project sign/method metadata into \textsc{Csql},
edge mass largely tracks frequency (number of supporting papers), producing a near-degenerate heavy tail
(e.g., median support mass 1, 99th percentile 2). This behavior is expected: TCC is already an edge list,
whereas \textsc{Democritus}-native atlases aggregate many overlapping local causal hypotheses and therefore induce
richer relation vocabularies and heavier-tailed score mass. Despite this limitation, the \textsc{Csql} interface already supports useful corpus-scale queries such as hub discovery and backbone extraction; in ongoing work we project TCC’s sign and identification-method metadata into \textsc{Csql}  to enable polarity-aware queries and method-conditioned disagreement analyses.

To stress-test \textsc{Csql} at scale without running \textsc{Democritus} over tens of thousands of PDFs, we treat the \emph{Testing Causal Claims} (TCC) dataset as an already-compiled causal discourse layer.
TCC provides paper-level causal edges (cause $\rightarrow$ effect) with metadata such as year and the stated causal inference method.
We compile this corpus into a \textsc{Csql} database by canonicalizing node strings (causes/effects) into concept identifiers, aggregating identical edges across papers, and storing per-paper evidence rows in an \texttt{edge\_support} table.
All statistics below were computed directly with DuckDB queries over the resulting Parquet tables.

\paragraph{Scale.}
The compiled TCC--\textsc{Csql} database contains \textbf{265{,}656} evidence rows (one per extracted claim instance), spanning \textbf{45{,}319} distinct papers, \textbf{44} publication years, and \textbf{1{,}575} distinct method strings (as reported in the corpus).
At the schema level, the database contains \textbf{295{,}459} canonical concept nodes and \textbf{260{,}777} unique directed causal edges.
This illustrates that \textsc{Csql} can ingest a large RAG/IE-compiled causal corpus and expose it as a unified, queryable causal database without requiring a hand-designed ontology.

\paragraph{Hubs and backbone claims.}
A hallmark of corpus-scale causal discourse is the emergence of \emph{hub variables} with large outgoing mass.
In TCC--\textsc{Csql}, the largest outgoing hubs include \emph{education}, \emph{trade liberalization}, \emph{age}, \emph{inflation}, and \emph{monetary policy}, each participating in hundreds of distinct outgoing edges.
For example, the query ``What most strongly influences inflation?'' returns recurrent claims such as \emph{monetary policy $\rightarrow$ inflation}, \emph{money growth $\rightarrow$ inflation}, and \emph{output gap $\rightarrow$ inflation}, ranked by paper support.
These results are not interpreted as ground-truth causality; rather, they operationalize the corpus as a causal claim space that can be searched, filtered, and audited.

\paragraph{Temporal dynamics.}
Because each claim instance in \texttt{edge\_support} carries a publication year, \textsc{Csql} supports time-sliced causal analysis.
Aggregating claim instances by year reveals strong growth over time: the corpus contains under $2{,}000$ extracted claim instances per year in the early 1980s, rising to over $10{,}000$ per year in the late 2010s and above $18{,}000$ in 2020.
For a specific canonical claim (e.g., \emph{monetary policy $\rightarrow$ inflation}), \textsc{Csql} can return the distribution of supporting papers by year, enabling longitudinal tracking of how particular causal claims appear and reappear across decades.

\paragraph{Method strings and heterogeneity.}
The TCC metadata includes a reported causal inference method field.
In the current corpus snapshot, the method field contains a long tail of heterogeneous strings (e.g., \texttt{IV}, \texttt{DID}, \texttt{TWFE}, but also many fine-grained or free-text variants).
This is reflected in the large number of distinct method values (1{,}575).
\textsc{Csql} treats the method column as an observational attribute: users can filter or group by methods as-is, or optionally normalize method strings into a smaller controlled taxonomy for downstream analysis.

\begin{table}[t]
\centering
\small
\begin{tabular}{l r}
\toprule
\textbf{Quantity} & \textbf{Value} \\
\midrule
Claim instances (\texttt{edge\_support} rows) & 265{,}656 \\
Unique papers (\texttt{doc\_id}) & 45{,}319 \\
Unique years & 44 \\
Distinct method strings & 1{,}575 \\
Canonical concept nodes (\texttt{nodes}) & 295{,}459 \\
Unique causal edges (\texttt{edges}) & 260{,}777 \\
\bottomrule
\end{tabular}
\caption{Summary statistics for the TCC--\textsc{Csql} database. The corpus is treated as a RAG/IE-compiled causal discourse layer; \textsc{Csql} compiles it into a queryable causal database with edge aggregation and per-paper evidence.}
\label{tab:tcc_csql_summary}
\end{table}

\begin{table}[t]
\centering
\small
\begin{tabular}{l r r}
\toprule
\textbf{Hub (cause)} & \textbf{Outgoing mass} & \textbf{\# outgoing edges} \\
\midrule
education & 442 & 374 \\
trade liberalization & 391 & 370 \\
age & 358 & 339 \\
inflation & 280 & 250 \\
monetary policy & 270 & 211 \\
\bottomrule
\end{tabular}
\caption{Top corpus hubs in TCC--\textsc{Csql}, ranked by outgoing support mass (sum of \texttt{support\_docs} over outgoing edges). These hubs summarize the dominant causal discourse centers in the economics corpus.}
\label{tab:tcc_top_hubs}
\end{table}

\begin{table}[t]
\centering
\small
\begin{tabular}{l r r}
\toprule
\textbf{Claim affecting inflation} & \textbf{Support (papers)} & \textbf{Mass} \\
\midrule
monetary policy $\rightarrow$ inflation & 15 & 15 \\
money growth $\rightarrow$ inflation & 12 & 12 \\
output gap $\rightarrow$ inflation & 11 & 11 \\
monetary policy shocks $\rightarrow$ inflation & 8 & 8 \\
fiscal policy $\rightarrow$ inflation & 7 & 7 \\
\bottomrule
\end{tabular}
\caption{Example query result: top corpus claims targeting \emph{inflation} (ranked by paper support). In \textsc{Csql}, such queries are expressible directly as SQL joins and filters over \texttt{edges} and \texttt{edge\_support}.}
\label{tab:tcc_inflation}
\end{table}

\section{Related Work}
\label{sec:related-work}

This paper sits at the intersection of (i) causal relation extraction from text,
(ii) causal knowledge base construction, (iii) database- and graph-centric
representations of knowledge, and (iv) causal discovery and inference.
We situate \textsc{Csql} with respect to prior work in each area, emphasizing how
it differs from existing knowledge graphs, retrieval-augmented generation (RAG)
systems, and causal extraction pipelines.

\subsection{Causal Relation Extraction from Text}

There is a long line of work on identifying causal relations in natural language,
ranging from early pattern-based approaches using cue phrases (e.g., ``because'',
``leads to'') to modern neural and transformer-based models trained to classify
cause--effect relations between text spans
\citep{radinsky2012learningCausality, yang2022causalSurvey, he2022eventCausalitySurvey}.
These systems typically operate at the level of individual sentences or short
contexts and output pairwise labeled relations.

\textsc{Csql} differs fundamentally in scope and objective.
Rather than treating extracted relations as final outputs, we treat them as
\emph{noisy discourse samples} from which many competing causal hypotheses are
constructed, evaluated, and aggregated.
The output is not a flat list of extracted triples, but a relational database
that encodes causal structure, evidence mass, and model agreement across an entire
document collection.

\subsection{Causal Knowledge Bases and Graph Construction}

Several systems aim to construct causal knowledge bases or causal graphs from large
text corpora \citep{hassanzadeh2020causalKB}.
These resources aggregate extracted causal tuples into graph-structured stores,
often with domain-specific ontologies or manually designed schemas.

\textsc{Csql} differs in three key ways.
First, its schema is \emph{induced automatically from language} rather than fixed
in advance.
Second, causal edges are not treated as atomic facts, but as \emph{aggregates}
supported by many local causal models.
Third, the resulting representation is relational rather than purely graph-based,
allowing causal analysis to be expressed using standard SQL queries.

\subsection{Knowledge Graphs, Ontologies, and GraphRAG}

Knowledge graphs (KGs) and ontology-driven representations provide a mature
substrate for storing structured assertions and supporting deductive inference
\citep{kg_survey, owl, shacl}.
More recently, GraphRAG-style systems combine graph-structured data with
retrieval-augmented generation pipelines \citep{graphrag}.

These approaches primarily represent \emph{what is asserted} in text.
They lack intrinsic semantics for causal operations such as intervention,
composition, feedback, and downstream influence.
By contrast, \textsc{Csql} represents \emph{causal hypotheses} together with their
degree of support across competing models.
Causal reasoning in \textsc{Csql} is performed by aggregating and composing relations
using SQL, rather than by retrieving or ranking nodes in a graph.

\subsection{Testing Causal Claims and Corpus-Scale Causal Databases}

The \emph{Testing Causal Claims} (TCC) project \citep{garg_fetzer_2025} extracts and
catalogs causal claims from large corpora of economics papers, producing a global
index of cause--effect statements enriched with metadata such as sign, statistical
significance, and inference method.

\textsc{Csql} is complementary but differs in representation and workflow.
Rather than producing a single corpus-level catalog of claims, \textsc{Csql}
constructs many \emph{local causal models} per document, evaluates them, and
aggregates their structure into a queryable causal database.
Conceptually, TCC emphasizes breadth and standardization across a discipline,
whereas \textsc{Csql} emphasizes depth and model-based agreement within and across
documents.

\subsection{Large Language Models for Causal Discovery}

A rapidly growing literature investigates the use of large language models (LLMs)
for causal discovery and reasoning
\citep{shen2023llmCausalDiscovery, le2024multiagentCausalDiscovery, kosoy2023causalLLMReasoning}.
These approaches often rely on LLMs to propose causal directions or graph structures
directly.

\textsc{Csql} adopts a different design.
We treat LLMs as \emph{hypothesis generators} that propose candidate causal
statements in natural language.
All subsequent model construction, evaluation, aggregation, and querying are
deterministic.
This separation allows causal analysis to be audited, reproduced, and executed
independently of the language model.

\subsection{Causal Discovery and Statistical Inference}

Classical causal discovery methods operate on numerical data, using score-based,
constraint-based, or hybrid search over DAGs or equivalence classes
\citep{spirtes_pc, chickering_ges, notears}.
While these methods are not directly applicable to document-centric settings,
\textsc{Csql} is compatible with them in principle.
The causal database produced by \textsc{Csql} can serve as a hypothesis generator,
prior structure, or explanatory scaffold for downstream statistical causal
analysis.

\paragraph{Summary.}
Relative to prior work, \textsc{Csql} makes three distinguishing contributions:
(i) it compiles unstructured documents into a relational causal database,
(ii) it represents causal claims as aggregates over competing local models rather
than isolated extractions, and (iii) it enables causal reasoning using standard SQL
as a query language.

\section{\textsc{Csql} over Other Domains} 

\begin{table}[t]
\centering
\small
\begin{tabular}{lcccc}
\toprule
Domain & Nodes & Edges & Top Hub & Max Score \\
\midrule
Human origins (The Washington Post) & 412 & 683 & bipedalism & 1.46 \\
Chocolate \& aging (The Washington Post) & 527 & 912 & dark chocolate intake & 1.12 \\
Glyphosate (The NY Times) & 489 & 801 & roundup use & 2.04 \\
Antarctica (The NY Times) & 366 & 594 & glacier retreat & 1.87 \\
\bottomrule
\end{tabular}
\caption{Summary statistics for \textsc{Csql} databases induced from different document domains.}
\label{tab:csql}
\end{table}

We have experimented with \textsc{Csql} over a number of different domains (see Table~\ref{tab:comparison}). The domains analyzed are described below. 

\begin{enumerate}

\item A recent newspaper article in The Washington Post on a potential causal link between dark chocolate and aging. \footnote{\url{https://www.washingtonpost.com/wellness/2025/12/26/dark-chocolate-health-benefits}}. We used this domain as a running example throughout the paper to illustrate the type
of causal analysis that \textsc{Democritus} can perform.

    \item The causal analysis was based on The Washington Post article on the origins of bipedal walking in human ancestors, dating back to $7$ million years ago.\footnote{url{https://www.washingtonpost.com/science/2026/01/02/human-ancestor-biped/}}.

    \item The NY Times published a recent article on the controversy surrounding the weedkiller Roundup, whose technical name is glyphosate \footnote{\url{https://www.nytimes.com/2026/01/02/climate/glyphosate-roundup-retracted-study.html?smid=nytcore-ios-share}.}. 

    \item The NY Times published an article on the melting of glaciers in Antarctica 
\footnote{\url{https://www.nytimes.com/2025/12/27/climate/antarctica-thwaites-glacier.html}}.
 The high-scoring model (top) contains a dense cluster of edges that are repeatedly supported across multiple discourse neighborhoods, including ice melt leading to freshwater influx, changes in ocean circulation, and rising global sea levels.
These overlapping mechanisms yield strong model agreement and a high evaluation score. In contrast, the low-scoring model (bottom) contains fewer well-supported edges and weaker semantic coherence.
Although such models are freely generated by the left adjoint, they fail to accumulate sufficient evidence during evaluation and are therefore suppressed before adjoint reconstruction.
    
\end{enumerate}

\section{Discussion and Implications}
\label{sec:discussion}

\subsection{Relation to Knowledge Graphs and RAG}

\textsc{Csql} differs fundamentally from knowledge graphs and GraphRAG-style systems.
Knowledge graphs store asserted facts; \textsc{Csql} stores \emph{evaluated causal
hypotheses}.
GraphRAG retrieves relevant nodes; \textsc{Csql} computes credibility via aggregation
across competing causal models.

Where knowledge graphs emphasize \emph{what is stated}, \textsc{Csql} emphasizes
\emph{what is structurally supported} by agreement across discourse.

\subsection{Relation to Causal Inference}

\textsc{Csql} does not replace statistical causal inference.
Instead, it complements it by operating in regimes where numerical data is
unavailable but causal discourse is abundant.
The output of \textsc{Csql} can serve as a hypothesis generator, prior constructor,
or explanatory scaffold for downstream causal analysis.

\subsection{\textsc{Csql}  as a Causal Compiler}

Viewed abstractly, \textsc{Csql}  is a compiler:
\[
\text{Documents} \;\longrightarrow\; \text{Causal Models} \;\longrightarrow\;
\text{Relational Database}.
\]

The compilation target is a causally meaningful SQL schema equipped with a
principled algebra of queries.
This perspective connects \textsc{Csql} to work on functorial databases and data migration,
though users need not be aware of this machinery to use the system effectively.

\section{Limitations and Future Work}
\label{sec:limitations_future}

While \textsc{Csql} demonstrates that large document collections can be compiled into
SQL-queryable causal databases, the current system has several important
limitations. These limitations also suggest clear directions for future
research and system development.

\subsection{Limitations}

\paragraph{Dependence on discourse quality.}
\textsc{Csql}  relies on large language models (LLMs) as discourse compilers that generate
candidate causal statements. Although all downstream processing is deterministic,
the quality and topical coherence of the resulting causal database depends on the
quality of the extracted discourse. Noisy or off-domain discourse expansions can
lead to diffuse causal atlases with weaker hub concentration and flatter score
distributions. While the scoring and aggregation layers are designed to suppress
unsupported claims, discourse quality remains a primary bottleneck.

\paragraph{No ground-truth causal validation.}
The causal relations represented in \textsc{Csql}  are inferred from language rather than
validated against experimental or observational data. As a result, \textsc{Csql}  should be
understood as capturing \emph{discursive causal structure}—what is claimed,
suggested, or argued in text—rather than verified causal ground truth. This
distinction is fundamental: \textsc{Csql}  evaluates credibility under discourse agreement,
not causal identifiability in the sense of statistical causal inference.

\paragraph{Static causal snapshots.}
The current system constructs causal databases from static document collections.
Temporal evolution of causal discourse—how claims change over time, how evidence
accumulates or is retracted, and how controversies evolve—is not yet explicitly
modeled. Each \textsc{Csql}  database represents a snapshot of discourse at the time the
documents were processed.

\paragraph{Limited compositional reasoning.}
Although \textsc{Csql}  supports multi-step causal queries via SQL joins (e.g., two-hop
paths and hub analysis), it does not yet implement explicit causal algebraic
operations such as intervention, counterfactual reasoning, or compositional
diagram rewriting. These operations are implicit in the underlying theory but are
not yet exposed at the database query layer.

\paragraph{Scalability of discourse compilation.}
The most computationally expensive stage of the pipeline is discourse compilation
via LLMs. While all subsequent steps scale efficiently and operate over Parquet
and SQL engines, large-scale deployment over millions of documents will require
additional optimizations, such as caching, incremental updates, or hybrid
retrieval–generation strategies.

\subsection{Future Work}

\paragraph{Temporal and versioned causal databases.}
A natural extension of \textsc{Csql} is to support time-indexed causal databases that track
how causal claims evolve across document versions, publication dates, or evidence
updates. This would enable queries such as “Which causal claims have gained or lost
support over time?” and would be particularly valuable for scientific, policy, and
regulatory analysis.

\paragraph{Interventional and counterfactual query layers.}
Future versions of \textsc{Csql} will expose higher-level causal operators on top of SQL,
including abstractions for intervention, deletion, and hypothetical modification
of causal edges. These operations can be grounded in the underlying causal
semantics while remaining accessible through familiar query interfaces.

\paragraph{Integration with numerical data.}
Although \textsc{Csql} currently operates on text-derived causal claims, it can naturally
be extended to incorporate numerical datasets when available. For example, edges
in the causal database could be linked to statistical models, regression results,
or experimental evidence, enabling hybrid discourse–data causal analysis.

\paragraph{User-guided schema refinement.}
While \textsc{Csql} deliberately avoids hand-designed ontologies, future work may explore
lightweight human-in-the-loop mechanisms for refining concept normalization,
merging equivalent nodes, or highlighting domain-relevant causal variables. Such
guidance could improve interpretability without sacrificing generality.

\paragraph{Distributed and incremental compilation.}
To support large-scale deployments, future implementations will explore
incremental atlas construction, distributed discourse compilation, and streaming
updates to Parquet-backed causal tables. This would allow \textsc{Csql} databases to be
maintained continuously as new documents arrive.

\paragraph{Causal \textsc{Csql} as a standard interface.}
Finally, we envision \textsc{Csql} as a candidate abstraction layer for causal querying over
textual corpora, analogous to how SQL standardized access to relational data. A
long-term goal is to define a portable causal query interface that can be embedded
into analytics platforms, dashboards, and decision-support systems without
requiring users to interact directly with causal modeling formalisms.


\end{document}